\documentclass[12pt]{article}
\usepackage{tikz}

\usepackage{CJKutf8}
\usepackage{amsmath,amssymb,graphicx,float}
\usepackage[unicode]{hyperref}
\usepackage{xcolor}
\usepackage{cite}
\usepackage{indentfirst}

\tikzset{
  treenode/.style = {shape=rectangle, rounded corners,
                     draw, align=center,
                     top color=white, bottom color=blue!20},
  root/.style     = {treenode, font=\Large, bottom color=blue!10},
  env/.style      = {treenode, font=\ttfamily\normalsize},
  dummy/.style    = {circle,draw}
}

\newcommand{\be}{\begin{equation}\begin{aligned}}
\newcommand{\bea}{\begin{eqnarray}}
\newcommand{\eea}{\end{eqnarray}}
\newcommand{\ba}{\begin{array}}
\newcommand{\ea}{\end{array}}
\newcommand{\ee}{\end{aligned}\end{equation}}

\newcommand{\al}{\alpha}

\newcommand{\tr}{\mbox{Tr}}

\begin{document}

\begin{titlepage}

\vspace*{5mm}%

\title{\textbf {Integrable matrix product state of ABJM theory from projecting method}}
	\author{Nan Bai$^{a}$\footnote{bainan@mailbox.gxnu.edu.cn}~,  Mao-Zhong Shao$^{a}$\footnote{mzshao@stu.gxnu.edu.cn}~
}
	\date{}
{\let\newpage\relax\maketitle}
	\maketitle
	\underline{}
	\vspace{-10mm}
	
	\begin{center}
		{\it
            $^{a}$ Department of Physics, Guangxi Normal University, \\Guilin 541004, China
		}
		\vspace{10mm}
	\end{center}

\begin{abstract}
In this paper we investigate the integrable boundary state in ABJM theory. We find an integrability condition for the two-site integrable matrix product state (MPS) similar to the KT-relation. We also construct a class of non-trivial MPSs from the projected K-matrices.
\end{abstract}
\end{titlepage}

\section{Introduction}
Integrable boundary state is a key object attracting much attention in the recent research for the exact results in quantum integrable systems, which was first proposed in the investigation of quantum quench models and defined as the state annihilated by the conserved charges with odd parities under space reflection\cite{Piroli:2017sei}. For the spin chain models, the integrability condition can be equivalently expressed by means of the transfer matrix and has a close relation with the boundary reflection equations\cite{Piroli:2017sei,Pozsgay:2018dzs}. As for the two-site integrable boundary state, a much more fundamental definition called KT-relation is further developed in \cite{Gombor:2021hmj}. One of the most prominent features of the integrable boundary state is the so-called pair structure which shows that the non-vanishing overlaps between the integrable boundary state and the Bethe state only occurs for the Bethe state with parity invariant root configurations and the resulted formulae for the overlaps have a factorized form given by the product of the polynomials of Q-functions and the ratio of the Gaudin-like determinants\cite{Jiang:2020sdw,Gombor:2020kgu,Kristjansen:2020vbe,Gombor:2021uxz,Gombor:2021hmj,Gombor:2024zru,Gombor:2024iix}. The integrable boundary state also finds its application in calculating the one-point function of the matrix product state (MPS) corresponding to various defect CFT configurations of the $\mathcal{N}=4$ SYM theory\cite{deLeeuw:2015hxa,Buhl-Mortensen:2015gfd,deLeeuw:2016umh,Buhl-Mortensen:2017ind,deLeeuw:2016ofj,DeLeeuw:2018cal}. Furthermore, it was also found that the three-point function of two determinant operators and a single trace operator can be exactly calculated when the determinant operators being the integrable MPS\cite{Jiang:2019zig,Jiang:2019xdz}.

The scope of the study of the integrable boundary state also extends to the ABJM theory\cite{Aharony:2008ug}, which is a three-dimensional Chern-Simons matter theory whose anomalous dimension matrix is mapped to an integrable alternating spin chain\cite{Minahan:2008hf,Bak:2008cp}. The exact correlation functions have been found in many cases where the integrable boundary states correspond to : determinant operators dual to giant gravitons\cite{Yang:2021hrl,Chen:2019kgc,Yang:2022dlk}; MPSs describing 1/2 BPS co-dimension one domain walls\cite{Gombor:2022aqj}; Wilson loop operators\cite{Jiang:2023cdm,Wu:2024uix}. In the present paper we consider the construction of the integrable MPSs of ABJM spin chain in a general way: We will use the projecting method \cite{Frahm:1998} to find a class of non-trivial operator-valued K-matrix solutions of the reflection equation and each projected K-matrix can be further used to construct a two-site invariant integrable MPS.

The paper is organized as follows: In the next section we review the integrable boundary state in ABJM theory. In section 3 we give a detailed classification of all the possible projected K-matrices. In the last section we give the conclusion and point out some future research directions.
\section{Integrable boundary state of ABJM spin chain}

In this section we will give a comprehensive review of the connection between the integrable boundary state and the reflection K-matrix for ABJM spin chain model. The demonstration given below basically follows the original papers \cite{Piroli:2017sei,Pozsgay:2018dzs} and  generalizes it to the open spin chain model with operator-valued K-matrix.

The ABJM spin chain model describes an alternating spin chain  with the adjacent sites carrying $\mathbf{4}$ and $\bar{\mathbf{4}}$ representations of $\rm{SU(4)}$ group respectively. The R-matrices can be defined either by the representation spaces they act upon, which results in four kinds of R-matrices $R_{12}(u),R_{1\bar{2}}(u),R_{\bar{1}2}(u)$ and $R_{\bar{1}\bar{2}}(u)$, or simply by introducing two different R-matrices given by
\begin{equation}
\begin{aligned}
R_{ij}(u)=u+\mathbb{P}_{ij},\quad \bar{R}_{ij}(u)=-u-2+\mathbb{K}_{ij},
\end{aligned}
\end{equation}
where $\mathbb{P}$ is the permutation operator and $\mathbb{K}$ is the partial transpose of $\mathbb{P}$: $\mathbb{K}_{ij}=\mathbb{P}_{ij}^{t_i}=\mathbb{P}_{ij}^{t_j}$. In order not to bring about any confusion, we will adopt the latter definitions throughout this paper. Then for a spin chain with 2L sites, we have the following two monodromy matrices,
\begin{equation}
\begin{aligned}
T(u)= R_{01}(u)\bar{R}_{02}(u)\cdots R_{0,2L-1}(u)\bar{R}_{0,2L}(u),\\
\overline{T}(u)= \bar{R}_{01}(u)R_{02}(u)\cdots \bar{R}_{0,2L-1}(u)R_{0,2L}(u),
\end{aligned}
\end{equation}
and also two related transfer matrices $\tau(u)=\tr_0T_0(u)$ and $\bar{\tau}(u)=\tr_0\overline{T}_0(u)$.
To study the open spin chain, we also need to introduce the boundary K-matrix which satisfies the following reflection equation (RE),
\begin{equation}\label{RE}
\begin{aligned}
R_{12}(u-v)K_1(u)R_{12}(u+v)K_2(v)=K_2(v)R_{12}(u+v)K_1(u)R_{12}(u-v).
\end{aligned}
\end{equation}
Note that if there exists additional internal degrees of freedom at the boundary, the matrix element of $K(u)$ will also be an operator of the internal space $V_{\rm{in}}$: $K^i_j\in {\rm{End}(V_{\rm{in}})}$. The component form of the above equation (\ref{RE}) is
\begin{equation}\label{cRE}
\begin{aligned}
R^{i,j}_{a_1,a_2}(u-v)K^{a_1}_{b_1}(u)R^{b_1,a_2}_{k,b_2}(u+v)K^{b_2}_l(v)\\=K^j_{a_2}(v)R^{i,a_2}_{a_1,b_2}(u+v)K^{a_1}_{b_1}(u)R^{b_1,b_2}_{k,l}(u-v),
\end{aligned}
\end{equation}
where the summation over repeated indices is implied. Multiplying the basis vector $|i,j,k,l\rangle$ of the tensor product space $V_1\otimes V_2\otimes V_3\otimes V_4$ to both sides of (\ref{cRE}), we find
\begin{equation}\label{cRE1}
\begin{aligned}
&R^{i,j}_{a_1,a_2}(u-v)K^{a_1}_{b_1}(u)R^{b_1,a_2}_{k,b_2}(u+v)K^{b_2}_l(v)|i,j,k,l\rangle\\
=&K^{a_1}_{b_1}(u)K^{b_2}_l(v)R^{b_1,a_2}_{k,b_2}(u+v)R_{12}(u-v)|a_1,a_2,k,l\rangle\\
=&K^{a_1}_{b_1}(u)K^{b_2}_l(v)R_{12}(u-v)R^{t_3}_{32}(u+v)|a_1,b_2,b_1,l\rangle,
\end{aligned}
\end{equation}
and
\begin{equation}\label{cRE2}
\begin{aligned}
&K^j_{a_2}(v)R^{i,a_2}_{a_1,b_2}(u+v)K^{a_1}_{b_1}(u)R^{b_1,b_2}_{k,l}(u-v)|i,j,k,l\rangle\\
=&K^j_{a_2}(v)K^{a_1}_{b_1}(u)R^{i,a_2}_{a_1,b_2}(u+v)R^{t_3t_4}_{34}(u-v)|i,j,b_1,b_2\rangle\\
=&K^j_{a_2}(v)K^{a_1}_{b_1}(u) R^{t_3t_4}_{34}(u-v) R^{t_4}_{14}(u+v)|a_1,j,b_1,a_2\rangle.
\end{aligned}
\end{equation}
If we define a two-site state as
\begin{equation}
\begin{aligned}
|\phi(u)\rangle=K^i_j(u)|i,j\rangle,
\end{aligned}
\end{equation}
then from Eqs (\ref{cRE}-\ref{cRE2}), we obtain
\begin{equation}\label{EQ1}
\begin{aligned}
&R_{12}(u-v)R^{t_3}_{32}(u+v)|\phi(u)\rangle_{13}\otimes |\phi(v)\rangle_{24}\\
=&R^{t_3t_4}_{34}(u-v) R^{t_4}_{14}(u+v) |\phi(v)\rangle_{24}\otimes |\phi(u)\rangle_{13}.
\end{aligned}
\end{equation}
Notice that both $R_{ij}(u)$ and $\bar{R}_{ij}(u)$ are P-symmetric, i.e. $R_{ij}(u)=R_{ji}(u)$, $\bar{R}_{ij}(u)=\bar{R}_{ji}(u)$, and also the relation
\begin{equation}
\begin{aligned}
\bar{R}_{ij}(u)=R^{t_{j}}_{ij}(-u-2),
\end{aligned}
\end{equation}
thus if we set $u=-1$ and then let $v=-u-1$ in (\ref{EQ1}) and also rename the space indices as: $2\rightarrow 0,\,3\rightarrow2,\,4\rightarrow0'$, we will finally have
\begin{equation}\label{BR}
\begin{aligned}
&R_{01}(u)\bar{R}_{02}(u)|\phi(-1)\rangle_{12}\otimes |\phi(-u-1)\rangle_{00'}\\
=&R_{0'2}(u)\bar{R}_{0'1}(u)|\phi(-u-1)\rangle_{00'}\otimes |\phi(-1)\rangle_{12}.
\end{aligned}
\end{equation}
Now we define the following tensor product state:
\begin{equation}\label{MPS1}
\begin{aligned}
|\Phi\rangle&=|\phi(-1)\rangle_{12}\otimes |\phi(-1)\rangle_{34} \otimes \cdots \otimes |\phi(-1)\rangle_{2L-1,2L}\\
&=K^{i_1}_{i_2}(-1)K^{i_3}_{i_4}(-1)\cdots K^{i_{2L-1}}_{i_{2L}}(-1)|i_1,i_2,\cdots,i_{2L-1},i_{2L}\rangle,
\end{aligned}
\end{equation}
which is a mixed-type state $|\Phi\rangle\in \mathcal{H}\otimes {\rm{End}(V_{in})}$, where $\mathcal{H}\cong (\mathbf{4}\otimes \bar{\mathbf{4}})^{\otimes L}$ is the Hilbert space of the whole spin chain. Tracing over the internal space $V_{\rm{in}}$, we obtain the two-site invariant matrix product state (MPS):
\begin{equation}\label{MPS2}
\begin{aligned}
|\Psi\rangle=\tr_{V_{\rm{in}}}|\Phi\rangle\,\in\,\mathcal{H}.
\end{aligned}
\end{equation}
By means of the basic relation (\ref{BR}), we can directly get
\begin{equation}
\begin{aligned}
T_0(u)|\Phi\rangle\otimes |\phi(-u-1)\rangle_{00'}=\left(\Pi\overline{T}_0(u)\Pi\right) |\phi(-u-1)\rangle_{00'}\otimes |\Phi\rangle,
\end{aligned}
\end{equation}
where $\Pi\in {\rm{End}(\mathcal{H})}$ is the space reflection operator of the spin chain. Now the trick here is that we first formulate the above relation in components on $V_0\otimes V_{0'}$ as
\begin{equation}
\begin{aligned}
&T^{\al}_{\lambda}(u)|\Phi\rangle \otimes K^{\lambda}_{\beta}(-u-1)|\al\rangle_{0}\otimes |\beta\rangle_{0'}\\
=&\left(\Pi\overline{T}^{\beta}_{\lambda}(u)\Pi\right)K^{\al}_{\lambda}(-u-1)|\al\rangle_{0}\otimes |\beta\rangle_{0'}\otimes |\Phi\rangle,
\end{aligned}
\end{equation}
then the component equality
\begin{equation}
\begin{aligned}
T^{\al}_{\lambda}(u)|\Phi\rangle K^{\lambda}_{\beta}(-u-1)=\left(\Pi\overline{T}^{\beta}_{\lambda}(u)\Pi\right)K^{\al}_{\lambda}(-u-1)|\Phi\rangle
\end{aligned}
\end{equation}
can be expressed on a single space $V_0$ as
\begin{equation}
\begin{aligned}
\left[T_0(u)\right]^{\al}_{\lambda}|\Phi\rangle \left[K_0(-u-1)\right]^{\lambda}_{\beta}
=\left[K_0(-u-1)\right]^{\al}_{\lambda}\Pi\left[\overline{T}_0(u)\right]^{\beta}_{\lambda}\Pi |\Phi\rangle,
\end{aligned}
\end{equation}
which leads to
\begin{equation}\label{Bf1}
\begin{aligned}
T_0(u)|\Phi\rangle K_0(-u-1)=K_0(-u-1)\left(\Pi\overline{T}^{t_0}_0(u)\Pi\right)|\Phi\rangle.
\end{aligned}
\end{equation}
The above relation is quite similar to the KT-relation for a general two-site integrable boundary state proposed in \cite{Gombor:2021hmj}. However there are two major differences: the spectral parameter dependence of $K_0$ and the form of the monodromy matrix on the right hand side of the equation \footnote{
As a comparison, we note that the KT-relation given in \cite{Gombor:2021hmj} has the form: 
\begin{align}\nonumber
K_0(u)\langle\Phi| T_0(u)=\langle\Phi| T_0(-u)K_0(u).
\end{align}}.

In the end, by multiplying $K^{-1}_0(-u-1)$ to the left on both sides of the above equation and then taking the trace over $V_0\otimes V_{\rm{in}}$, we obtain
\begin{equation}
\begin{aligned}
\tau(u)|\Psi\rangle=\Pi \bar{\tau}(u) \Pi |\Psi\rangle,
\end{aligned}
\end{equation}
which is the final form of the integrability condition for integrable boundary state in ABJM theory used in the literatures \cite{Gombor:2020kgu,Yang:2021hrl,Gombor:2022aqj}.

\section{Integrable MPS from the projected K-matrix}
In this section we will investigate the possible operator-valued K-matrix solutions by the projecting method proposed in \cite{Frahm:1998}.  The projecting K-matrix for ABJM spin chain can be obtained as follows: Given a c-number solution $K_1(u)\in {\rm{End}}(V_1)$ of RE (\ref{RE}), one can construct the following two operator-valued solutions $K_{12}(u)$ and $\bar{K}_{12}(u)$ on $V_1\otimes V_2$ by means of the two R-matrices $R_{12}(u)$ and $\bar{R}_{12}(u)$ respectively as:
\begin{equation}\label{RegK}
\begin{aligned}
K_{12}(u)=R_{12}(u+c)K_1(u)R_{12}(u-c),\\
\bar{K}_{12}(u)=\bar{R}_{12}(u+\bar{c})K_1(u)\bar{R}_{12}(u-\bar{c}),
\end{aligned}
\end{equation}
where $V_2$ is the additional inner space at the boundary and $c$($\bar{c}$) are two arbitrary constants. We will call them the regular solutions. However such solutions are trivial since they simply represent adding one more quantum site to the bulk of the original spin chain \cite{Sklyanin:1988yz}.

To find the non-trivial solutions, we introduce the projection operator $Q$: $Q^2=Q,\,Q\in {\rm{End}{(V_2)}}$.  It can be easily shown that if either of the regular solution $K_{12}(u)$ or $\bar{K}_{12}(u)$ in (\ref{RegK}) satisfies the following projection condition:
\begin{equation}\label{proj}
\begin{aligned}
Q_2 K_{12}(u)(\mathbb{I}-Q_2)=0\quad \mbox{or} \quad Q_2 \bar{K}_{12}(u)(\mathbb{I}-Q_2)=0,
\end{aligned}
\end{equation}
then $Q_2 K_{12}(u) Q_2$ or $Q_2 \bar{K}_{12}(u) Q_2$ becomes the non-trivial operator-valued solution of RE and will be called the projected K-matrix.

 As the initial input, the c-number solution of RE (\ref{RE}) takes the general form $K(u)=\xi \mathbb{I}+u \mathbb{E}$ where $\xi$ is an arbitrary constant and $\mathbb{E}$ can be chosen from either of the following two classes (details can be found in the paper \cite{Arnaudon:2004sd}):
 \begin{itemize}
   \item [(1)] $\mathbb{E}$ is diagonal and $\mathbb{E}^2=\mathbb{I}$;
   \item [(2)] $\mathbb{E}$ is strictly triangular and $\mathbb{E}^2=0$.
 \end{itemize}

So a natural question here is: based on the above two kinds of c-number solutions and the two possible constructions of regular solutions (\ref{RegK}), can we find the suitable projector $Q$ satisfying the projection condition (\ref{proj}) and then obtain the corresponding projected K-matrix solution ? In the following we will answer this question by giving all possible projectors for each case and thus obtain a complete classification of the integrable MPS from the projected K-matrix.

\subsection{Projected K-matrix from $K_{12}(u)$}
We expand $K_{12}(u)$ by the order of spectral parameter $u$ as:
\begin{equation}
\begin{aligned}
K_{12}(u)=&u^3 \mathbb{E}_1+u^2\left(\xi \mathbb{I}+\mathbb{E}_1 \mathbb{P}_{12}+\mathbb{P}_{12} \mathbb{E}_1\right)\\
+&u\left(2\xi \mathbb{P}_{12}-c^2 \mathbb{E}_1+c \mathbb{E}_1 \mathbb{P}_{12}-c \mathbb{P}_{12} \mathbb{E}_1+\mathbb{E}_2\right)+\xi\left(1-c^2\right)\mathbb{I}.
\end{aligned}
\end{equation}
From the projection condition $Q_2 K_{12}(u) \left(\mathbb{I}-Q_2\right)=0$, we observe that the coefficient in front of $u^3$ gives,
\begin{equation}
\begin{aligned}
Q_2 \mathbb{E}_1 \left(\mathbb{I}-Q_2\right)=\mathbb{E}_1\left(Q_2-Q^2_2\right)=0,
\end{aligned}
\end{equation}
which means $Q$ must be a projector. Therefore we could begin with an arbitrary operator $Q$ with 16 unknowns $\{q_{ij},i,j=1,\cdots,4\}$ (not necessary a projector),
\begin{equation}
Q=\left(
\begin{array}{cccc}
q_{11}&q_{12}&q_{13}&q_{14}\\
q_{21}&q_{22}&q_{23}&q_{24}\\
q_{31}&q_{32}&q_{33}&q_{34}\\
q_{41}&q_{42}&q_{43}&q_{44}\\
\end{array}
\right),
\end{equation}
and then impose the projection condition to $Q$, then all the surviving $Q$s are automatically being the projection operators we search for.

\subsubsection{Case A: $K_{12}(u)$ with the first class of the c-number solution}
There are further two concrete choices for the first class of c-number solutions which lead to non-equivalent projectors. We now discuss them separately.

If the c-number solution is
  \begin{equation}\label{cs1}
  K(u)=\left(
  \begin{array}{cccc}
  \xi+u&0&0&0\\
  0&\xi-u&0&0\\
  0&0&\xi-u&0\\
  0&0&0&\xi-u
  \end{array}\right),
  \end{equation}
  we find:
  \begin{itemize}
    \item[(a)] For $c\neq \pm\xi$ , we only have two trivial solutions : $Q=0$ and $Q=\mathbb{I}$;
    \item[(b)] For $c=\xi \neq 0$, the non-trivial projector is
  \begin{equation}
  Q=\left(
  \begin{array}{cccc}
  1&0&0&0\\
  \alpha&0&0&0\\
  \beta&0&0&0\\
  \gamma&0&0&0
  \end{array}\right), \quad \forall \al,\beta,\gamma\in \mathbb{C};
  \end{equation}
  \item[(c)] For $c=-\xi\neq 0$, the non-trivial projector is
  \begin{equation}
  Q=\left(
  \begin{array}{cccc}
  0&\al&\beta&\gamma\\
  0&1&0&0\\
  0&0&1&0\\
  0&0&0&1
  \end{array}\right), \quad \forall \al,\beta,\gamma\in \mathbb{C};
  \end{equation}
  \item[(d)] For $c=\xi=0$, we have both of the above two non-trivial solutions:
  \begin{equation}
  Q_1=\left(
  \begin{array}{cccc}
  0&\al&\beta&\gamma\\
  0&1&0&0\\
  0&0&1&0\\
  0&0&0&1
  \end{array}\right),\quad
  Q_2=\left(
  \begin{array}{cccc}
  0&\al&\beta&\gamma\\
  0&1&0&0\\
  0&0&1&0\\
  0&0&0&1
  \end{array}\right), \quad \forall \al,\beta,\gamma\in \mathbb{C}.
  \end{equation}
  \end{itemize}

If the c-number solution is
  \begin{equation}\label{cs2}
  K(u)=\left(
  \begin{array}{cccc}
  \xi+u&0&0&0\\
  0&\xi+u&0&0\\
  0&0&\xi-u&0\\
  0&0&0&\xi-u
  \end{array}\right),
  \end{equation}
  we find:
  \begin{itemize}
    \item[(a)] For $c\neq\pm\xi$, we only have the trivial solutions: $Q=0$ and $Q=\mathbb{I}$;
    \item[(b)] For $c=\xi\neq 0$, the non-trivial solution is,
    \begin{equation}
  Q=\left(
  \begin{array}{cccc}
  1&0&0&0\\
  0&1&0&0\\
  \al&\beta&0&0\\
  \gamma&\delta&0&0
  \end{array}\right), \quad \forall \al,\beta,\gamma,\delta \in \mathbb{C};
  \end{equation}
    \item[(c)] For  $c=-\xi\neq 0$, the non-trivial solution is,
  \begin{equation}
  Q=\left(
  \begin{array}{cccc}
  0&0&\al&\beta\\
  0&0&\gamma&\delta\\
  0&0&1&0\\
  0&0&0&1
  \end{array}\right), \quad \forall \al,\beta,\gamma,\delta \in \mathbb{C};
  \end{equation}
  \item[(d)] For $c=\xi=0$, we have two non-trivial solutions:
  \begin{equation}
  Q_1=\left(
  \begin{array}{cccc}
  1&0&0&0\\
  0&1&0&0\\
  \al&\beta&0&0\\
  \gamma&\delta&0&0
  \end{array}\right), \quad
  Q_2=\left(
  \begin{array}{cccc}
  0&0&\al&\beta\\
  0&0&\gamma&\delta\\
  0&0&1&0\\
  0&0&0&1
  \end{array}\right), \quad \forall \al,\beta,\gamma,\delta \in \mathbb{C}.
  \end{equation}
  \end{itemize}

\subsubsection{Case B:  $K_{12}(u)$ with the second class of the c-number solution}\label{subs1}
For the second class of the c-number solution, since $\mathbb{E}$ is a nilpotent matrix of degree 2 and strictly triangular, we could choose the following two basic forms for $\mathbb{E}$:
\begin{equation}
E=\left(
\begin{array}{cccc}
0&1&&\\
0&0&&\\
&&0&1\\
&&0&0
\end{array}
\right) \quad {\mbox{or}} \quad
E=\left(
\begin{array}{cccc}
0&1&&\\
0&0&&\\
&&0&0\\
&&0&0
\end{array}
\right),
\end{equation}
and thus $K(u)$ will take the corresponding block diagonal forms shown below.

If the c-number solution is
\begin{equation}\label{cs3}
K(u)=\left(
\begin{array}{cccc}
\xi& u&&\\
0&\xi&&\\
&&\xi&u\\
&&0&\xi
\end{array}
\right),
\end{equation}
we find:
\begin{itemize}
  \item[(a)] For $\xi\neq 0$, there are only two trivial solutions: $Q=0$ and $Q=\mathbb{I}$;
  \item[(b)] For $\xi=0$, besides the trivial solutions, we also have another non-trivial solution:
   \begin{equation}
Q=\left(
\begin{array}{cccc}
0&\alpha&0&\beta\\
0&1&0&0\\
0&\gamma&0&\delta\\
0&0&0&1
\end{array}
\right),\quad \forall \alpha,\beta,\gamma,\delta\in \mathbb{C};
\end{equation}
\end{itemize}
If the c-number solution is
\begin{equation}\label{cs4}
K(u)=\left(
\begin{array}{cccc}
\xi& u&&\\
0&\xi&&\\
&&\xi&0\\
&&0&\xi
\end{array}
\right),
\end{equation}
we find:
\begin{itemize}
  \item[(a)] For $\xi\neq 0$, only trivial solutions exist;
  \item[(b)] For $\xi=0$, the non-trivial solutions are quite abundant as listed below:
  \begin{equation}\label{QPATH1}
  \begin{aligned}
  &Q_1=\left(
  \begin{array}{cccc}
  0&\al&\beta&\gamma\\
  0&1&0&0\\
  0&0&1&0\\
  0&0&0&1
  \end{array}
  \right),\quad
  Q_2=\left(
  \begin{array}{cccc}
  0&\delta&\al\beta&\al\\
  0&1&0&0\\
  0&\gamma&0&0\\
  0&-\gamma\beta&\beta&1
  \end{array}
  \right),\\
  &Q_3=\left(
  \begin{array}{cccc}
  0&\alpha&\beta&0\\
  0&1&0&0\\
  0&0&1&0\\
  0&\gamma&\delta&0
  \end{array}
  \right),\quad
  Q_4=\left(
  \begin{array}{cccc}
  0&\alpha&0&0\\
  0&1&0&0\\
  0&\beta&0&0\\
  0&\gamma&0&0
  \end{array}
  \right),\\
  &Q_5=\left(
  \begin{array}{cccc}
  0&\alpha&\beta&0\\
  0&1&0&0\\
  0&\gamma&0&\delta\\
  0&0&0&1
  \end{array}
  \right),\quad
  Q_6=\left(
  \begin{array}{cccc}
  0&\delta&\alpha&\alpha\beta\\
  0&1&0&0\\
  0&-\beta\gamma&1&\beta\\
  0&\gamma&0&0
  \end{array}
  \right),\\
  &Q_7=\left(
  \begin{array}{cccc}
  0&\mu&\delta&{\delta\gamma}/{\alpha}\\
  0&1&0&0\\
  0&-{\beta\gamma}/{\alpha}&1-\gamma&{(1-\gamma)\gamma}/{\alpha}\\
  0&\beta&\alpha&\gamma
  \end{array}
  \right),\,(\alpha\neq0),
  \end{aligned}
  \end{equation}
  where all the free parameters above are arbitrary complex numbers. In the appendix \ref{AP1} we give the concrete paths for the generation of these projectors.
\end{itemize}
From the c-number solutions and the corresponding $Q$-projectors, we find a large class of non-trivial K-matrices: $\mathcal{K}_{12}(u)= Q_2K_{12}(u)Q_2$, and then the integrable MPSs can be obtained from these projected K-matrices by the constructions in (\ref{MPS1}) and (\ref{MPS2}).

\subsection{Projected K-matrix from $\bar{K}_{12}(u)$}
By the same argument as in the case above, we only need to solve the projection condition $Q_2 \bar{K}_{12}(u)\left(\mathbb{I}-Q_2\right)=0$ to find the appropriate $Q$s.
\subsubsection{Case A:  $\bar{K}_{12}(u)$ with the first class of c-number solution}
For the c-number solution in (\ref{cs1}), we find:
\begin{itemize}
    \item[(a)] For the case $\bar{c}\neq\pm(\xi+1)$, we only have two trivial solutions;
    \item[(b)] For the case $\bar{c}=\xi+1,\bar{c}\neq 0$, the non-trivial solution is,
    \begin{equation}
  Q=\left(
  \begin{array}{cccc}
  0&\al&\beta&\gamma\\
  0&1&0&0\\
  0&0&1&0\\
  0&0&0&1
  \end{array}\right), \quad \forall \al,\beta,\gamma \in \mathbb{C};
  \end{equation}
    \item[(c)] For $\bar{c}=-\xi-1,\bar{c}\neq 0$, the non-trivial solution is
    \begin{equation}
  Q=\left(
  \begin{array}{cccc}
  1&0&0&0\\
  \al&0&0&0\\
  \beta&0&0&0\\
  \gamma&0&0&0
  \end{array}\right). \quad \forall \al,\beta,\gamma \in \mathbb{C};
  \end{equation}
   \item[(d)] For $\xi=-1,\bar{c}=0$, there are two non-trivial solutions:
    \begin{equation}
  Q_1=\left(
  \begin{array}{cccc}
  0&\al&\beta&\gamma\\
  0&1&0&0\\
  0&0&1&0\\
  0&0&0&1
  \end{array}\right), \quad
  Q_2=\left(
  \begin{array}{cccc}
  1&0&0&0\\
  \al&0&0&0\\
  \beta&0&0&0\\
  \gamma&0&0&0
  \end{array}\right), \quad \forall \al,\beta,\gamma \in \mathbb{C}.
  \end{equation}
  \end{itemize}

  For the c-number solution in (\ref{cs2}), we find:
  \begin{itemize}
    \item[(a)] For $\bar{c}\neq \pm \xi$, there are only two trivial solutions;
    \item[(b)] For $\bar{c}=\xi, \bar{c}\neq 0$, the non-trivial solution is
    \begin{equation}
  Q=\left(
  \begin{array}{cccc}
  0&0&\al&\beta\\
  0&0&\gamma&\delta\\
  0&0&1&0\\
  0&0&0&1
  \end{array}\right), \quad \forall \al,\beta,\gamma,\delta \in \mathbb{C};
  \end{equation}
    \item[(c)] For $\bar{c}=-\xi, \bar{c}\neq 0$, the non-trivial solution is
    \begin{equation}
  Q=\left(
  \begin{array}{cccc}
  1&0&0&0\\
  0&1&0&0\\
  \al&\beta&0&0\\
  \gamma&\delta&0&0
  \end{array}\right), \quad \forall \al,\beta,\gamma,\delta \in \mathbb{C};
  \end{equation}
    \item[(d)] For $\bar{c}=\xi=0$, we have both of the above two non-trivial solutions:
    \begin{equation}
    Q_1=\left(
  \begin{array}{cccc}
  0&0&\al&\beta\\
  0&0&\gamma&\delta\\
  0&0&1&0\\
  0&0&0&1
  \end{array}\right),\quad
  Q_2=\left(
  \begin{array}{cccc}
  1&0&0&0\\
  0&1&0&0\\
  \al&\beta&0&0\\
  \gamma&\delta&0&0
  \end{array}\right), \quad \forall \al,\beta,\gamma,\delta \in \mathbb{C}.
    \end{equation}
  \end{itemize}
\subsubsection{Case B:  $\bar{K}_{12}(u)$ with the second class of c-number solution}
For the c-number solution in (\ref{cs3}), we find:
\begin{itemize}
  \item[(a)] For $\xi\neq 0$, there are only two trivial solutions;
  \item[(b)] For $\xi=0$, the non-trivial solution is
    \begin{equation}
  Q=\left(
  \begin{array}{cccc}
  1&0&0&0\\
  \alpha&0&\beta&0\\
  0&0&1&0\\
  \gamma&0&\delta&0
  \end{array}\right), \quad \forall \alpha,\beta,\gamma,\delta \in \mathbb{C}.
  \end{equation}
\end{itemize}
For the c-number solution in (\ref{cs4}), we find:
\begin{itemize}
  \item[(a)] For $\xi\neq 0$, there are only two trivial solutions;
  \item[(b)] For $\xi=0$, there exists seven non-trivial projectors given below:
  \begin{equation}\label{QPATH2}
  \begin{aligned}
  &Q_1=\left(
  \begin{array}{cccc}
  1&0&0&0\\
  \alpha&0&\beta&0\\
  0&0&1&0\\
  \gamma&0&\delta&0
  \end{array}
  \right),\quad
  Q_2=\left(
  \begin{array}{cccc}
  1&0&0&0\\
  \alpha&0&\beta&\gamma\\
  0&0&1&0\\
  0&0&0&1
  \end{array}
  \right),\\
  &Q_3=\left(
  \begin{array}{cccc}
  1&0&0&0\\
  \alpha&0&0&0\\
  \beta&0&0&0\\
  \gamma&0&0&0
  \end{array}
  \right),\quad
  Q_4=\left(
  \begin{array}{cccc}
  1&0&0&0\\
  \alpha&0&0&\beta\\
  \gamma&0&0&\delta\\
  0&0&0&1
  \end{array}
  \right),\\
  &Q_5=\left(
  \begin{array}{cccc}
  1&0&0&0\\
  \delta&0&\alpha&\alpha\beta\\
  -\beta\gamma&0&1&\beta\\
  \gamma&0&0&0
  \end{array}
  \right),\quad
  Q_6=\left(
  \begin{array}{cccc}
  1&0&0&0\\
  \delta&0&\alpha\beta&\alpha\\
  \gamma&0&0&0\\
  -\beta\gamma&0&\beta&1
  \end{array}
  \right),\\
  &Q_7=\left(
  \begin{array}{cccc}
  1&0&0&0\\
  \mu&0&\delta&\gamma\delta/\alpha\\
  -\beta\gamma/\alpha&0&1-\gamma&(1-\gamma)\gamma/\alpha\\
  \beta&0&\alpha&\gamma
  \end{array}
  \right),\,(\alpha\neq 0).
  \end{aligned}
  \end{equation}
  The routes to obtain these $Q$s are shown in the appendix \ref{AP1}.
  \end{itemize}
Similarly, the corresponding integrable MPSs are obtained from the projected K-matrices constructed by the projection operators given above.

\section{Conclusion and discussion}
In this paper we studied the integrable boundary state in ABJM theory from two aspects: We first reconsidered the relation between the two-site integrable MPS and the operator-valued solution of reflection equation for the ABJM spin chain. In this process, we found an integrability condition for the MPS which is quite similar to the KT-relation proposed in \cite{Gombor:2021hmj}; We also constructed a class of integrable MPSs by the projecting method and gave a complete classification of the projected K-matrices by finding all possible non-trivial projection operators.

For future research, an immediate direction is to investigate the exact overlaps for the integrable MPSs obtained from projected K-matrices. Besides, in parallel to the studies given in the present work, we could also consider the operator-valued solutions of the twisted reflection equation which represents the soliton non-preserving boundary conditions \cite{Pozsgay:2018dzs,Delius:1998he,Doikou:2000yw}, and then construct the related integrable MPSs.

\begin{appendix}
\section{The flow chart for the generation of the projection operators}\label{AP1}
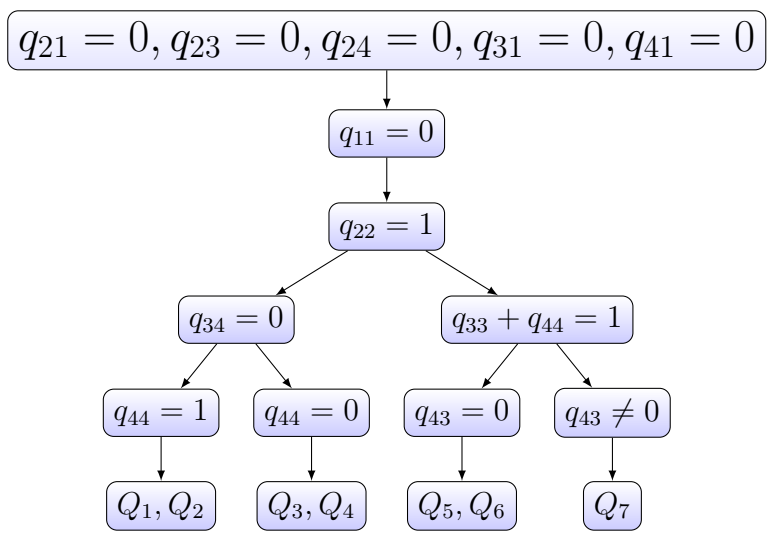
\begin{figure}[htb]
  \centering
\begin{tikzpicture}
  [
    level distance          = 3em,
    edge from parent/.style = {draw, -latex},
    every node/.style       = {font=\footnotesize},
    sloped
  ]
  \node [root] {$q_{21}=0,q_{23}=0,q_{24}=0,q_{31}=0,q_{41}=0$}[sibling distance =4cm]
    child { node [env] {$q_{11}=0$}[sibling distance =2cm]
            child { node [env] {$q_{22}=1$} [sibling distance=4cm]
                   child { node [env] {$q_{34}=0$}[sibling distance =2cm]
                            child{node[env]{$q_{44}=1$}
                                  child { node[env]{$Q_1,Q_2$}
                                        }
                                  }
                            child{node[env]{$q_{44}=0$}
                                  child { node[env]{$Q_3,Q_4$}
                                         }
                                  }
                         }
                   child { node [env] {$q_{33}+q_{44}=1$}[sibling distance =2cm]
                            child { node[env]{$q_{43}=0$}
                                    child { node[env]{$Q_5,Q_6$}
                                          }
                                    }
                            child { node[env]{$q_{43}\neq0$}
                                     child { node[env]{$Q_7$}}
                                   }
                         }
                  }
          };
\end{tikzpicture}
 \caption{The generation paths of the  projectors in (\ref{QPATH1}).}\label{fig1}
\end{figure}

Here we give the concrete generation paths for the two classes of the projectors given in the main text:

In the $K_{12}(u)$ case, for the c-number solution (\ref{cs4}) with $\xi=0$, from the projection condition (\ref{proj}), we find the following matrix elements of $Q$ are identically zero,
\begin{equation}
\begin{aligned}
q_{21}=q_{23}=q_{24}=q_{31}=q_{41}=0.
\end{aligned}
\end{equation}
Then the way to obtain the seven projectors in (\ref{QPATH1}) is shown in figure (\ref{fig1}).

In the $\bar{K}_{12}(u)$ case, for the c-number solution in (\ref{cs4}) with $\xi=0$ , the projection condition (\ref{proj}) gives
\begin{equation}
\begin{aligned}
q_{12}=q_{13}=q_{14}=q_{32}=q_{42}=0,
\end{aligned}
\end{equation}
and the projectors in (\ref{QPATH2}) can be obtained as shown below in figure (\ref{fig2}).
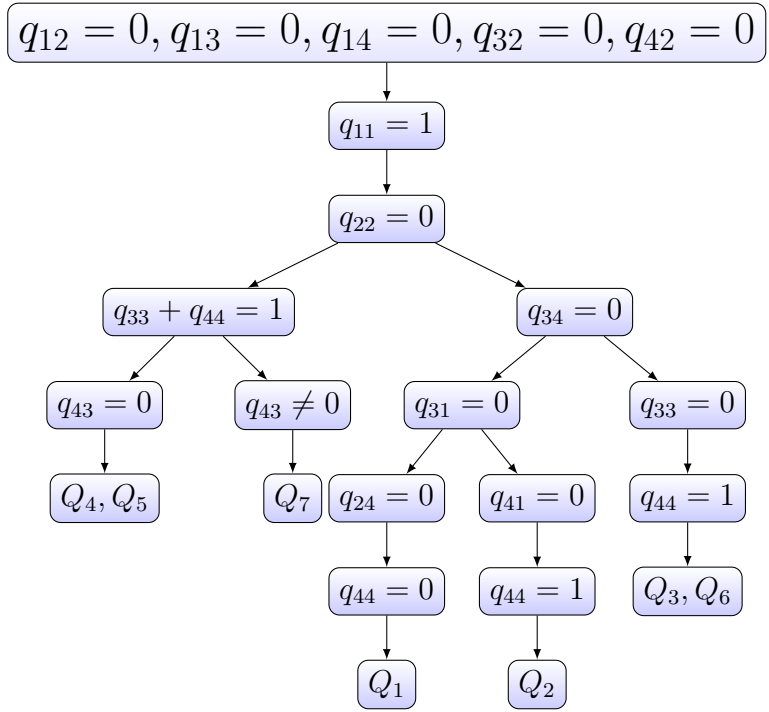
\begin{figure}[h]
  \centering
\begin{tikzpicture}
  [
    grow                    = down,
    level distance          = 3em,
    edge from parent/.style = {draw, -latex},
    every node/.style       = {font=\footnotesize},
    sloped,
  ]
  \node [root] {$q_{12}=0,q_{13}=0,q_{14}=0,q_{32}=0,q_{42}=0$}[sibling distance =4.5cm]
     child { node [env]{$q_{11}=1$}
            child { node[env]{$q_{22}=0$}[sibling distance =5cm]
                    child { node[env]{$q_{33}+q_{44}=1$}[sibling distance =2.5cm]
                            child { node [env] {$q_{43}=0$}
                                    child { node [env] {$Q_4,Q_5$}
                                           }
                                  }
                             child { node [env] {$q_{43}\neq0$}
                                     child {node [env] {$Q_{7}$}
                                           }
                                   }
                          }
                      child { node[env]{$q_{34}=0$}[sibling distance =3cm]
                            child { node [env] {$q_{31}=0$}[sibling distance =2cm]
                                child { node [env] {$q_{24}=0$}[sibling distance =1cm]
                                       child { node [env] {$q_{44}=0$}
                                              child { node [env] {$Q_1$}
                                                    }
                                              }
                                       }
                                child { node [env] {$q_{41}=0$}[sibling distance =1cm]
                                      child { node [env] {$q_{44}=1$}
                                                child { node [env] {$Q_2$}
                                                      }
                                             }
                                      }
                                  }
                            child { node [env] {$q_{33}=0$}[sibling distance =1cm]
                                    child { node [env] {$q_{44}=1$}
                                            child { node [env] {$Q_3,Q_6$}
                                                  }
                                          }
                                  }
                          }
                  }
            };
\end{tikzpicture}
 \caption{The generation paths of the projectors in (\ref{QPATH2}).}\label{fig2}
\end{figure}
\end{appendix}

\end{document}